\documentstyle[12pt]{article}
\def\vect#1#2{\setbox0=\hbox{$#1$}   \def\cpy{\copy0\kern-\wd0}
              \def\column{\raise.025em\cpy \raise.0125em\cpy \cpy}
              \def\back{\kern-#2em}  \def\forw{\kern+#2em}
        \hbox{\raise.0250em\cpy      \back\back\column
              \forw\column           \forw\forw\column
              \forw\column           \back\back\box0  }}

\def\t{\vect{\tau}{0.015}}

\begin{document}

\title{ Constraining the Strongly-Coupled Standard Model
        Including a $W'$ Isotriplet }

\author{Eric Sather$^{1,\>2,\>3}$ \\
Stanford Linear Accelerator Center \\
Stanford University \\
Stanford, CA 94309 \\
\\ and \\
\\ Witold Skiba$^{2,\>3}$ \\
Center for Theoretical Physics \\
Massachusetts Institute of Technology \\
Cambridge, MA 02139}
\addtocounter{footnote}{1}\footnotetext{This work is supported in
part by the U.S. Department of Energy under
contract DE-AC03-76SF00515.\hfill}
\addtocounter{footnote}{1}\footnotetext{This work is supported in
part by the U.S. Department of Energy under
contract DE-AC02-76ER03069.\hfill}
\addtocounter{footnote}{1}\footnotetext{This work is supported in
part by the National Science Foundation under Grant
No.\ PHY89-04035.\hfill}

\date{}

\maketitle

\vspace{-5.5in}
\rightline{ \begin{tabular}{l}
SLAC-PUB-6657 \\
MIT-CTP-2409  \\
NSF-ITP-95-09  \\
February 1995  \\
T/E           \\
\end{tabular} }
\vspace{ 4.2in}

\begin{center}{Submitted to: {\it Physical Review D}}\end{center}

\begin{abstract}
We consider the Strongly-Coupled Standard Model (Abbott-Farhi model)
including an isotriplet of $W'$ vector bosons.  First we calculate
the corrections to the low-energy theory, which can be effectively
summarized in terms of the parameters $S$, $T$ and $U.$  Then we use
high-precision electroweak measurements to constrain the mass and
couplings of the $W'$.  The $W'$ couplings are restricted to be
unnaturally small, and we conclude that this model is no longer
compelling as a theory of the electroweak interactions.
\end{abstract}

\vfill

\thispagestyle{empty}

\newpage
\setcounter{page}{1}

\section{Introduction}

With the ever-increasing precision of electroweak measurements,
viable theories of physics beyond the Standard Model have become
somewhat scarce.  In this paper, we reexamine a candidate theory
which promises a rich spectrum of resonances and other
strong-interaction phenomena beyond the weak scale: the
Strongly-Coupled Standard Model (SCSM), or Abbott-Farhi model
\cite{af}.

The SCSM is based on an underlying lagrangian identical in form to
the Standard-Model lagrangian.  However, the parameters of the
gauge-Higgs sector are adjusted so that the Higgs field does not
spontaneously break the SU(2)$_L$ gauge symmetry.  Instead the
SU(2)$_L$ interactions become confining, and the observed particle
spectrum consists of SU(2)$_L$ singlets.  Nevertheless, given
dynamical assumptions such as unbroken chiral symmetry, the
low-energy theory of the SCSM looks very much like the
spontaneously-broken Standard Model. The striking similarity of the
confining and spontaneously broken phases of the theory exemplifies
the concept of ``complementarity.'' The exciting possibility that
nature might in fact be described by a confining version of the
Standard Model, which predicts the discovery of new particles and
strong interactions at future colliders, motivates the study of the
SCSM.

Of course, if the SCSM really is the theory of the weak interactions,
evidence for particle compositeness must eventually emerge.  The
effective theory of the SCSM must deviate from the renormalizable
Standard-Model lagrangian: resonances and higher-dimensional
interactions should appear.  In this work we ask: Are the deviations
expected in the SCSM allowed by current experimental constraints? We
will attempt to answer this question by studying a test case in which
we introduce an isotriplet of $W'$ vector bosons into the effective
theory.  We calculate the corrections to Standard-Model predictions
which result from including the $W'$ bosons. Then we use
high-precision electroweak data to constrain these corrections, and
thereby bound the allowed region of the $W'$ mass and couplings.

In an earlier analysis of experimental constraints on the SCSM, Korpa
and Ryzak \cite{kr}\ added both an $W'$ isotriplet and an isoscalar
vector boson to the usual Standard-Model particle content. They
concluded that experiment could accommodate these new particles
without severely restricting their masses and couplings.  Since then
many new electroweak observables have been measured, and the accuracy
of earlier measurements has greatly increased. Here we exploit this
new data to find much stronger constraints on the model. In
particular, the precise measurement of the $Z$ mass permits a new
approach to parameterizing the corrections to Standard-Model
predictions in the SCSM.

In the original formulation of the model \cite{af}, the $W$ and $Z$
were not expected to have the masses predicted by the Standard Model;
however, it was known that the Standard-Model values could be
recovered by invoking vector dominance \cite{hs}. Specifically, they
follow from the assumption that the $W$-pole graph saturates the
isovector electromagnetic form factor of the composite fermions.
When the $W$ and $Z$ were later discovered with masses near the
Standard-Model values, it became necessary to add to the SCSM the
assumption that vector dominance holds at least approximately
\cite{cfj}. In our analysis we find that the $W$ must saturate this
form factor to within a few percent.  We consider this unnatural and
conclude that, in its present form, the SCSM is no longer a candidate
for a theory of the electroweak interactions.

In the next Section we review the effective theory for the SCSM in a
limit where non-Standard particle content and higher-dimensional
interactions are absent, and show that it reduces to the Standard
Model in this limit. In Section~3 we introduce a $W'$ isotriplet and
discuss the resulting modification of vector-dominance relations.  In
Section~4 we calculate the $W'$-induced corrections to electroweak
observables, which we summarize in terms of contributions to $S$,
$T$ and $U$ in Section~5.  We then use high-precision electroweak
measurements in Section~6 to determine the allowed region of the $W'$
mass and couplings, and present our conclusions in the final Section.

\section{Review of the SCSM}

The fundamental insight which underlies the SCSM is that the particle
spectrum and interactions in the strong-coupling version of the
Standard Model could closely resemble those of the familiar
spontaneously-broken Standard Model. This is an example of
complementarity: there is no phase transition between confinement and
spontaneous symmetry breaking in a gauge-Higgs theory with a Higgs in
the fundamental representation \cite{comp}.  In this Section we review
how the effective theory of the SCSM approximates the ordinary
Standard Model (given certain dynamical assumptions).  We closely
follow the presentation and notation of Claudson, Farhi and Jaffe
\cite{cfj}. In the next Section we will begin our discussion of the
deviations from Standard-Model predictions which appear when a $W'$
isotriplet is added.

The SCSM is based on an underlying lagrangian which has the same form
as the Standard-Model lagrangian.  However, the parameters of the
theory are adjusted so that the SU(2)$_L$ interactions are not
spontaneously broken, and instead become confining at low energies.
All the observed particles are then SU(2)$_L$ singlets.

Consider the potential for the fundamental scalar field,
\begin{equation}
V(\Omega)={\lambda\over2}
(\mathrel{\rm tr}\Omega^{\dag}\Omega-2v^2)^2\ ,
\label{omegapot}
\end{equation}
where
\begin{equation}
\Omega=\left(\matrix{\phi&\tilde\phi\cr}\right)
  =\left(\matrix{\phi_1&-\phi_2^*\cr
   \phi_2&\phi_1^*\cr}\right)\ .
\label{omdef}
\end{equation}
Note that $\Omega^{\dag}\Omega=|\phi|^2{\bf1}.$ By expressing the
potential in terms of $\Omega$, we make explicit the invariance of
the potential under the custodial SU(2)$_W$ symmetry, defined by
$\Omega\to\Omega h$ for $h\in$SU(2)$_W$. This symmetry is an
invariance of the full lagrangian when the hypercharge and Yukawa
couplings of the fermions are turned off.

The scale dependence of the SU(2)$_L$ gauge coupling is characterized
by a scale parameter $\Lambda_2$, analogous to $\Lambda_{\rm QCD}$.
This scale parameter and the constant $v^2,$ which appears in the
scalar-field potential in Eq.~(\ref{omegapot}), together control
whether the SU(2)$_L$ interactions are confining or spontaneously
broken. The SU(2)$_L$ gauge symmetry will not be spontaneously broken
if $v^2<0$ or if $v^2\ll\Lambda_2^2$, in which case the gauge
interactions get strong at energies well above $v^2$ and prevent
spontaneous symmetry breaking.  The fundamental fields which carry
SU(2)$_L$ charge will then be confined into SU(2)$_L$ singlets. These
can be classified using the custodial symmetry, SU(2)$_W$.

For example, the elementary left-handed fermions $\psi_L^a$ (where
$a=1,\ldots,12$ labels the SU(2)$_L$ doublet) bind with the scalar
particles $\phi$ to form composite left-handed fermions,
\begin{equation}
F_L^a=\Omega^{\dag}\psi_L^a=\left(\matrix{\phi^{*\alpha}
   {\psi_L^a}_\alpha\cr
   \phi_\alpha\epsilon^{\alpha\beta}{\psi_L^a}_\beta\cr}\right),
\end{equation}
which transform as SU(2)$_W$ doublets. Here $\alpha$ and $\beta$ are
SU(2)$_L$ indices, which are contracted so that the $F_L^a$ are
SU(2)$_L$ singlets. The hypercharge of a composite fermion is the sum
of the elementary fermion and scalar hypercharges, $y^a+\tau^3/2.$
This is simply the electric charge $Q^a$ of the fermion, which
implies that the hypercharge U(1) in the SCSM is actually
electromagnetism.

{}From the scalar fields alone we can form a composite Higgs field,
$H={1\over2}\mathrel{\rm tr}(\Omega^{\dag}\Omega)$, which is an
SU(2)$_W$ singlet. We can also form an SU(2)$_W$ triplet of vector
bosons, with interpolating field ${\bf W}_\mu=\mathrel{\rm tr}
(\Omega^{\dag}D_\mu\Omega\t).$  In these examples we can see the
crucial role played by the custodial symmetry in organizing the
composite particles into multiplets analogous to the familiar
SU(2)$_L$ multiplets of the Standard Model. (We will later see how
this symmetry also ensures that the interactions of the composite
particles have the standard form.)

Of course, in addition to the particles that are contained in the
Standard Model, experience with the Strong Interactions leads us to
expect in the SCSM a rich spectrum of bound states, including excited
$W'$ bosons, leptoquarks and so on. Since these particles have yet to
be observed, we must assume that these exotic states are considerably
more massive than the left-handed fermions and the $W$ bosons.

Claudson, Farhi, and Jaffe enumerated three dynamical assumptions
concerning the confining SU(2)$_L$ sector of the theory, which must hold
if the SCSM is to describe the observed electroweak phenomena
\cite{cfj}:
\begin{itemize}
\item[(i)] The approximate SU(12) chiral symmetry which relates the
12 SU(2)$_L$ fermion doublets is not spontaneously broken by a
condensation of left-handed fermions (i.e.,
$\langle\psi_L^a\psi_L^b\rangle=0$).  This chiral symmetry then
protects the composite left-handed fermions $F_L^a$ from acquiring
large masses.  (If this chiral symmetry were broken, there would be
light Goldstone bosons consisting of two left-handed fermions, and
the composite fermions would be heavy, as their analogs are in QCD.)
 \item[(ii)] The $W$ vector bosons are much lighter than the
typical mass scale in the theory (e.g., $\Lambda_2$), and in
particular, the $W$ and $Z$ are much lighter than their recurrences,
the $W'$ and $Z'$.
 \item[(iii)] The effective coupling of the $W$
bosons to left-handed fermions is small ($\bar{g}\approx0.66$) even
while the underlying theory is strongly-coupled. \end{itemize}

With these assumptions we can write down the low-energy effective
lagrangian for the SCSM.  Interactions with dimension greater than
four should be suppressed by the characteristic mass scale,
$\Lambda_2,$ which by assumption (ii) is much larger than $M_W.$  As
long as we work at energies no higher than the $Z$ mass, we should be
able to omit these higher-dimensional operators from the effective
theory. Then the most general SU(2)$_W$-symmetric effective
lagrangian involving the composite fermion and vector-boson fields is
\begin{equation}
{\cal L}_{\rm eff}^0=i\bar{F}_{La}\partial\!\!\!/
   F_L^a-{1\over4}{\bf W}^{\mu\nu}\cdot {\bf W}_{\mu\nu}
   +{1\over2}M_W^2{\bf W}^\mu\cdot{\bf W}_\mu
   +\bar{g}{\bf W}_\mu\cdot{\bf j}_L^\mu+\ldots,
\label{lzero}
\end{equation}
where the $W$ self-couplings have not been listed. Here ${\bf
j}_L^\mu={1\over2} \bar{F}_{La} \t\gamma^\mu F_L^a,$ and ${\bf
W}_{\mu\nu}=\partial_\mu {\bf W}_\nu-\partial_\nu {\bf W}_\mu$
Electromagnetism, which breaks the custodial symmetry, can then be
added by minimal substitution of the vector potential, $a_\mu,$ and
insertion of the field-strength, ${\cal F}_{\mu\nu}= \partial_\mu
a_\nu-\partial_\nu a_\mu$:
\begin{equation}
{\cal L}_{\rm eff}={\cal L}_{\rm eff}^0+i\bar{\psi}_{Rb}\partial\!\!\!/
   \psi_R^b
   +e a_\mu j_{\rm em}^\mu-{1\over4}{\cal F}^{\mu\nu}{\cal F}_{\mu\nu}
   - {k\over2}{\cal F}^{\mu\nu}W^3_{\mu\nu}+\cdots.
\label{lem}
\end{equation}
Again cubic and quartic vector-boson interactions have not been
listed. Here $j_{\rm em}^\mu$ is the contribution of the fermions to
the electromagnetic current. If we now assume vector dominance, so
that the isovector electromagnetic form factor of the $F_L^a$ is
saturated by the $W$ boson, then as will be discussed in the next
Section, we find that the strength $k$ of the photon-$W^3$
mixing\footnote{This mixing is analogous to the familiar case of
photon-$\rho$ mixing.} is given in terms of the U(1) coupling and the
$W\bar{F}_L F_L$ coupling as $k=e/\bar{g}.$ Diagonalizing the
quadratic terms in the lagrangian (\ref{lem}) which involve the
neutral vector bosons, we find the propagating fields
\begin{eqnarray}
A_\mu&=&a_\mu+k W^3_\mu \> , \nonumber\\
Z_\mu&=&(1-k^2)^{1/2}W^3_\mu \> ,
\label{prop}
\end{eqnarray}
which couple to the neutral currents as
\begin{equation}
{\cal L}_{\rm NC}=e A \cdot j_{\rm em}
  + Z\cdot {\bar{g}\over\sqrt{1-k^2}}\left(j_L^3
  -{ek\over\bar{g}}j_{\rm em}\right)\> .
\end{equation}
Hence the value of $\sin^2\theta$ that would be measured in
low-energy neutrino scattering is $\sin^2\theta=ek/\bar{g}=k^2,$
where we have used the vector-dominance result $k=e/\bar{g}.$  This
implies that $\bar{g}=e/\sin\theta,$ which leads to the standard
prediction for the mass of the $W$: $M_W^2= \pi\alpha/ (\sqrt{2}G_F
\sin^2\theta).$ In the above diagonalization process one additionally
finds a $Z$ mass of $M_Z = M_W/\sqrt{1-k^2}.$ Applying the
vector-dominance result again, we recover the Standard-Model relation
\begin{equation}
{M_W^2\over{M_Z}^2\cos^2\theta}=1\>.
\label{massrel}
\end{equation}
We see therefore how the additional assumption of vector dominance
leads to the Standard-Model predictions for the masses of both the
$W$ and the $Z.$ A vector-dominance analysis of the electromagnetic
form factors of the $W$ shows that the cubic and quartic
self-couplings of the $W$ are those of an SU(2) gauge theory with
coupling $\bar{g},$ and that the corresponding couplings of the
propagating fields, $A$ and $Z$, are just those of the Standard
Model \cite{cfj}.

\section{Including a $W'$ Isotriplet}

We have just seen that with certain dynamical assumptions, and
invoking vector dominance, the effective theory of the SCSM
approximates the Standard Model. We now begin our analysis of the
corrections to the effective theory that result when we introduce an
isotriplet of $W'$ vector bosons. A $W'$ should arise in this model
as a radial excitation of the $W$, analogous to the $\rho'$ in QCD.
Because there is no evidence yet for deviations from the Standard
Model, the $W'$ must be considerably more massive than the $W$ and/or
less strongly coupled. We will therefore treat the inclusion of the
$W'$ as a perturbation of the Standard Model.

Of course, we could include other non-Standard particles in the
theory. Alternatively, we could include in the lagrangian all
operators up to some dimension which are consistent with the
symmetries of the theory. However, as we will soon show, the $W'$ is
the degree of freedom which corresponds to relaxing the assumption
that the $W$ saturates the isovector electromagnetic form factor.
Adding a $W'$ thus allows us to consider corrections to the effective
theory due to new particle content, and also to study deviations from
exact vector dominance. At the same time, including a $W'$ isotriplet
adds only three new parameters to the low-energy effective theory,
and therefore it is possible to significantly constrain the theory.
The lagrangian terms for the $W'$ are similar in form to those for
the $W,$ Eqs.~(\ref{lzero}) and (\ref{lem}):
\begin{eqnarray}
{\cal L}_{\rm eff}^{W'}&=&-{1\over4}{{\bf W}'}^{\mu\nu} \cdot
   {{\bf W}'}_{\mu\nu}
   + {1\over2}{M_W'}^2\>{{\bf W}'}^\mu\cdot{{\bf W}'}_\mu
   +\bar{g}'{{\bf W}'}_\mu\cdot{\bf j}_L^\mu \nonumber\\
   &&\quad -{k'\over2}{\cal F}^{\mu\nu}{{W'}^3}_{\mu\nu}+\cdots.
   \label{lem'}
\end{eqnarray}
In constructing ${\cal L}_{\rm eff}^{W'}$ we have proceeded much as
before in arriving at Eqs.~(\ref{lzero}) and (\ref{lem}).  We have
first constructed the most general $SU(2)_W$-symmetric lagrangian.
Next we have diagonalized the lagrangian to eliminate terms which mix
the $W$ and $W'$ bosons. Finally we have included electromagnetism,
which leads to mixing of the photon with $W^3$ and ${W'}^3.$
For later purposes, we note that by substituting ${\bf W}'\to-{\bf
W}'$ we can reverse the signs of both $\bar{g}'$ and $k',$ showing
that only the relative sign of these two couplings is meaningful.

\subsection{Vector Dominance}

We now show how the $W'$ parameterizes the deviation from
vector-meson dominance: Consider the isovector electromagnetic form
factor of the composite fermions, $F_V(q^2)$, defined through
\begin{equation}
e\langle\bar{F}_{La},\,F_L^a|J^\mu_{\rm em}|0\rangle=
e\bar{U}_L\gamma^\mu y^aV_L+e\, F_V(q^2)\,
\bar{U}_L\gamma^\mu{\tau^3\over2}V_L\>.
\label{form}
\end{equation}
Here $J^\mu_{\rm em}$ is the total electromagnetic current, which
includes terms linear in $W^3$ and $W'^3.$ The sum of the
contributions to the form factor (see Fig.~1) from direct coupling of
the current to the left-handed fermions and from the $W^3$- and
${W'}^3$-pole diagrams is
\begin{equation}
eF_V(q^2)=e-k\bar{g} {q^2\over q^2-M_W^2}
  - k'\bar{g}' {q^2\over q^2-M_W'^2}\>.
\end{equation}
Because the $F_L$ are composite and the confining SU(2)$_L$
interactions are asymptotically free, $F_V(q^2)\to0$ as
$|q^2|\to\infty$.  Then, assuming that $F_V(q^2)$ is saturated by the
$W^3$ and ${W'}^3$ poles (i.e., there are no other contributions to
the form factor, such as from a $W''$), we conclude that
\begin{equation}
e=k\bar{g}+k'\bar{g}'=k\bar{g}(1+\kappa\gamma)\>,
\label{vectdom}
\end{equation}
where we have introduced the ratios of $W$ and $W'$ couplings
\begin{equation}
\kappa\equiv k'/k\>,\quad \gamma\equiv\bar{g}'/\bar{g}\>.
\end{equation}
Had we assumed strict vector dominance so that only the $W$-pole
diagram contributed, we would have found the result $e=k\bar{g}$,
which leads to the Standard-Model lagrangian as was shown in
Section~2.  By including the $W',$ however, we depart from
exact vector dominance. The $W'$ contributes a fraction
$k'\bar{g}'/e\approx\kappa\gamma$ to the saturation of $F_V(q^2)$, so
that $\kappa\gamma$ measures the degree of the departure.

For the model to approximately reproduce the Standard Model, the $W$
must nearly saturate this form factor. Here we have further assumed
that the $W'$ contribution to the saturation, though of necessity
much smaller than that of the $W$, is nevertheless more important
than contributions from higher-lying resonances, which have been
ignored. In essence, we are claiming that the $W'$ can be viewed as a
stand-in for all the resonances beyond the $W$ which contribute to
$F_V(q^2)$.  The quantity $\kappa\gamma$ accordingly represents the
combined contribution of these resonances to the saturation of this
form factor.

\subsection{The Physical Neutral Vector Bosons}

The mixing of the photon with the neutral $W^3$ and ${W'}^3$ bosons
introduces off-diagonal terms into the free (quadratic) part of the
lagrangian.  We need to diagonalize the free lagrangian to find the
physical photon, $Z$ and $Z'$ fields. In the previous Section we gave
expressions for the physical photon and $Z$ fields that diagonalize
the free part of lagrangian in the absence of the $W'$ bosons. These
expressions, and also the result for the $Z$ mass, are modified by
the mixing of the photon with ${W'}^3.$ For the masses of the $Z$ and
$Z'$ we find
\begin{eqnarray}
M_Z^2&\approx&{M_W^2\over c^2}\left[1-{s^4\over c^2(c^2-\mu)}
   \mu\kappa^2 \right]\ ,\nonumber\\
{M_{Z'}}^2&\approx&{M_{W'}}^2\left[
   {1+{s^4\over c^4}\mu\kappa^2 \over 1-{s^2\over c^2}\kappa^2}
   +{s^4\over c^2(c^2-\mu)}\mu\kappa^2 \right]\ .
\label{mzsqr}
\end{eqnarray}
Here $s\equiv\sin\theta\equiv k$, $c\equiv\cos\theta$, and $\mu$ is
the ratio of the $W$ and $W'$ squared masses:
\begin{equation}
\mu\equiv M_W^2/{M_{W'}}^2\ .
\label{mdef}
\end{equation}
Terms containing extra factors of $\mu\kappa^2$ have been omitted.
Note that we must restrict $\kappa$ to the interval
$|\kappa|<\cot\theta$, since otherwise the $Z'$ would have a negative
squared mass and be a tachyon.  The neutral vector-boson fields $a$,
$W^3$ and ${W'}^3$ are given in terms of the physical fields $A,$ $Z$
and $Z'$ as
\begin{eqnarray}
a&=&A-s(W^3+\kappa{W'}^3)\ ,\nonumber\\
    W^3&\approx&\left[1-\left(1-{\mu\over2c^2}\right)
    \left({s^2\over c^2-\mu}\right)^2
    \mu\kappa^2\right]{Z\over c}
    + {s^2\over c^2-\mu} \mu\kappa
    {Z'\over\sqrt{1-{s^2\over c^2}\kappa^2}}\ ,\nonumber\\
    W'^3&\approx&-{s^2\over c^2-\mu} \mu\kappa{Z\over c}
    + {Z'\over\sqrt{1-{s^2\over c^2}\kappa^2}}\ .
\label{neutralphys}
\end{eqnarray}
Corrections to the coefficients of $Z$ and $Z'$ are suppressed by
additional factors of $\mu\kappa^2$.  If we substitute these
expressions into the interaction terms that couple $a$, $W$, and $W'$
to the fermions, we find the couplings of the $Z$ and $Z'$ to
left-handed and electromagnetic currents.

These results for the $Z$ and $Z'$ masses (\ref{mzsqr}) and the
expressions for $W^3$ and ${W'}^3$ in terms of $Z$ and $Z'$
(\ref{neutralphys}) were calculated by expanding in powers
$s^2\mu\kappa/(c^2-\mu)\ll1,$ which is assumed to be small.  This
assumption reflects our intuition that the $W'$ should be heavier
than the $W$ ($\mu\ll1)$ and should also mix more weakly with the
photon ($\kappa\ll1$).

Note that all of the above corrections to Standard-Model relations
--- the vector dominance result for the electromagnetic coupling
(\ref{vectdom}) and the mass and couplings of the $W$ and $Z,$
(\ref{mzsqr}) and (\ref{neutralphys}) --- contain factors of at least
two of the $W'$ parameters $\kappa$, $\gamma$ and $\mu$.  The same is
true of corrections to four-fermi interactions mediated by $W'$
exchange, which are of order $\mu\gamma^2$.  Hence if any two of the
$W'$ parameters vanish, the effective low-energy theory reduces to
that of the Standard Model, leaving the remaining $W'$ parameter
completely unconstrained.  This means that we will be unable to
obtain constraints on any individual parameter independent of the
other parameters.  We will either have to fix one of the parameters
and then constrain the other two, or else constrain products of the
parameters, e.g., the product $\kappa\gamma$.

As mentioned above, Korpa and Ryzak in their earlier analysis of SCSM
constraints considered the SCSM with not only an isotriplet of $W'$
vector bosons but also with isoscalar vector bosons which are bound
states of a left-handed fermion and a left-handed antifermion:
$(V_\mu)_a^b \sim \bar{\psi}_{La}\gamma_\mu\psi_L^b.$  Assuming that
those $V$ bosons which are color octets, and thus mix with gluons,
saturate the (isoscalar) color form factor of the composite,
left-handed quarks, they were able to place a very stringent bound on
the mass of the isoscalar bosons ($m_V > 700\,{\rm GeV}$), so that the
$V$ bosons would be just as massive as the rest of the non-Standard
resonances. This called into question the assumption of vector
dominance of the isoscalar form factors. Of course, vector dominance
of the {\it isoscalar} channel is not a necessary ingredient in the
SCSM.  By contrast, vector dominance of the {\it isovector} channel
must at least approximately hold in order to account for the $W$ and
$Z$ masses. By including a $W'$ we can determine to what accuracy
vector dominance must be maintained in the isovector channel in order
to retain agreement with electroweak data.

\section{Corrections to Standard-Model Predictions}

We have seen that the SCSM as formulated here reduces to the Standard
Model if the $W'$ is absent or if it has infinite mass and vanishing
couplings.  And, of course, the Standard Model is in impressive
agreement with experiment.  It is therefore logical to treat the SCSM
with a $W'$ as a perturbation of the Standard Model and calculate
the corrections to SM predictions due to the $W'$.

\subsection{Corrections to the Mass and Couplings of the $W$}

The corrections induced by the $W'$ are not simply given by the sum
of the new graphs that include $W'$s.  The three quantities $\alpha$,
$G_F$ and $M_Z$ are known to very high accuracy, and their values
cannot change when the $W'$ bosons are added to the theory. In the
Standard Model, their values determine the masses and couplings of
the vector bosons.  However, as the $W'$ parameters are turned on,
the $W$ parameters must deviate from their Standard-Model values if
$\alpha$, $G_F$ and $M_Z$ are to remain fixed. There are then two
ways in which the $W'$ modifies the effective theory: (1)~$W'$
exchange induces new effective (four-fermi) interactions, {\it and}
(2)~the $W$ mass and couplings depart from their Standard-Model
values in order to preserve the values of $\alpha$, $G_F$, and $M_Z$.

Let us first compute the deviations of the $W$ mass and couplings
from their Standard-Model values by working at tree level. We define
$M_{W0}$, $\bar{g}_0$ and $k_0$ as the mass and couplings of the
$W$ when the $W'$ is absent, and we define $M_W,$ $\bar{g}$ and $k$
as the mass and couplings when the $W'$ is included.  We compute $e,$
$G_F$ and $M_Z$ (at tree level) in the Standard Model when the $W'$
is absent, and then in the SCSM, with a $W'$ in the theory.
Combining these results we have
\begin{eqnarray}
e&=&\ \ k_0\bar{g}_0\ \ =\ k\bar{g}(1+\kappa\gamma)\ ,\nonumber\\
    4\sqrt{2}G_F&=&{\bar{g}_0^2\over{M_{W0}}^2}\ =\
    {\bar{g}^2\over{M_W}^2}+{\bar{g}'^2\over{M_{W'}}^2}\ =\
       {\bar{g}^2\over{M_W}^2}(1+\mu\gamma^2)\ ,\nonumber\\
    M_Z^2&=&{{M_{W0}}^2\over c_0^2}\ \approx\
       {M_W^2\over c^2}
       \left[1-{s^4\over c^2(c^2-\mu)}\mu\kappa^2\right]\ .
\label{fixed}
\end{eqnarray}
Here $c_0\equiv\cos\theta_0=\sqrt{1-k_0^2},$ and as before, $s=k$ and
$c=\sqrt{1-k^2}.$  The expressions for $e=\sqrt{4\pi\alpha}$ and
$M_Z$ come from the previous Section, Eqs.~(\ref{vectdom}) and
(\ref{mzsqr}).  The result for $G_F$ is simply the sum of $W$ and
$W'$ exchange. These formulas can be used to express the deviations
of the $W$ mass and couplings from their Standard-Model values in
terms of ${M_{W'}}$, $\bar{g}'$ and $k'$ (or, equivalently, in terms
of $\mu$, $\kappa$ and $\gamma$). Let $\delta M_W=M_W-{M_{W0}}$,
$\delta\bar{g}=\bar{g}-\bar{g}_0$ and $\delta k=k-k_0$.  Then
\begin{eqnarray}
{\delta M_W\over M_W}\approx&&{1\over2}{1\over c^2-s^2}
  \left[2s^2\kappa\gamma-s^2\mu\gamma^2+{s^4\over c^2-\mu}
  \mu\kappa^2\right]\ ,\nonumber\\
{\delta\bar{g}\over\bar{g}}\approx&&{1\over2}{1\over c^2-s^2}
  \left[2s^2\kappa\gamma-c^2\mu\gamma^2+{s^4\over c^2-\mu}
  \mu\kappa^2\right]\ ,\nonumber\\
{\delta k\over  k}\approx&-&{1\over2}{1\over c^2-s^2}
  \left[2c^2\kappa\gamma-c^2\mu\gamma^2+{s^4\over c^2-\mu}
  \mu\kappa^2\right]\ .
\label{shifts}
\end{eqnarray}
Terms containing more powers of $\kappa\gamma$, $\mu\gamma^2$ or
$\mu\kappa^2$ have been omitted. The corrections to Standard-Model
predictions we will find, such as all those found above, will in
general be linear in these quantities, and so they will be forced to
be small by our constraint analysis.
In Eq.~(\ref{shifts}) where $s$ and $c$ multiply small quantities
like $\kappa\gamma$, we could just as well use $s_0$ and $c_0,$ since
the expressions would be unchanged within the accuracy to which we
are working.  Here and in the following, wherever the choice of
$\sin\theta_W$ is immaterial, we will use $\sin\theta=k$ and write
simply $s$ (and $c$) for brevity.  It should be understood however,
that another convention would do just as well.

These results have been obtained at tree level.  Of course, in
addition to the $W'$-induced corrections to the effective theory,
there are also radiative corrections.  Radiative corrections to the
small $W'$-induced corrections are negligible, comparable to two-loop
corrections in the Standard Model.  Hence the effective theory is
well approximated by adding the $W'$-induced corrections calculated
here at tree level to the Standard-Model effective theory calculated
to one-loop accuracy.  In particular, the mass and couplings of the
$W$ in the SCSM are obtained from their renormalized Standard-Model
values by simply adding the deviations $\delta M_W,$ $\delta\bar{g}$
and $\delta k$ calculated above.

\subsection{Neutral-Current Interactions at the $Z$ Pole}

Having calculated the correction to $M_W,$ we need to compute the
corrections to the neutral-current interactions in order to obtain
the remaining constraints on the $W'$ parameters. (Charged-current
interactions are precisely constrained only at low energy, and there
they are completely fixed by the value of $G_F$.) Above we presented
expressions for the fields $a$, $W$ and $W'$ in terms of the physical
neutral vector bosons $A$, $Z$ and $Z'$ (\ref{neutralphys}).  Using
those results we can write down the coupling of the physical bosons
to the fermion currents ${j_L^3}$ and $j_{\rm em}$ (again, we only
need to calculate the $W'$-induced corrections at tree-level):
\begin{eqnarray}
{\cal L}_{\rm NC}&=&ea\cdot j_{\rm em} + \bar{g} W^3\cdot{j_L^3} +
   \bar{g}'{W'}^3\cdot{j_L^3}\nonumber\\
&\approx&eA\cdot j_{\rm em} \nonumber\\
&&+\bar{g}\left[1-{1\over2}{s^4\over c^2}{2c^2-\mu\over(c^2-\mu)^2}
  \mu\kappa^2-{s^2\over c^2-\mu}\mu\kappa\gamma\right]\nonumber\\
&&\ \times{Z\over c}\cdot\left[{j_L^3}-{ek\over\bar{g}}
   \left(1+{s^2\over c^2-\mu}
  \mu\kappa(\gamma-\kappa)\right)j_{\rm em}\right]\nonumber\\
&&+\bar{g} Z'\cdot\left[\left(\gamma+{s^2\over c^2-\mu}\kappa\right)
   {j_L^3} -{ek\over\bar{g}}\left(1+{s^2\over c^2-\mu}\right)\kappa
   j_{\rm em}\right]
\label{lnc}\\ \nonumber\\
&\approx&eA\cdot j_{\rm em} \nonumber\\
&&+\bar{g}_0(1+\zeta){Z\over c_0}\cdot\left[{j_L^3}-s_0^2(1+\delta_Z)
   j_{\rm em}\right]
\nonumber\\
&&+\bar{g} Z'\cdot\left[\sqrt{-\zeta\over\mu}({j_L^3}-s^2 j_{\rm em})
   -s^2(\kappa-\gamma) j_{\rm em}\right].
\label{lncsimple}
\end{eqnarray}
{}From the coupling of the $Z$ in (\ref{lncsimple}), we see that
corrections to observables measured at the $Z$ pole are summarized by
the quantities $\delta_Z$ and $\zeta,$ which are given by
\begin{eqnarray}
\delta_Z&=&{\delta\sin^2\theta\over\sin^2\theta}\Bigg|_{Z\ {\rm pole}}
  =-{c^2\over c^2-s^2} \left[{s^2\over c^2-\mu}\mu\kappa^2
  +{1-(1+2s^2)\mu\over c^2-\mu}\kappa\gamma-\mu\gamma^2\right]\ ,\\
\zeta&=&-\mu\left(\gamma+{s^2\over c^2-\mu}\kappa\right)^2\ .
\label{szdef}
\end{eqnarray}
$\delta_Z$ is the fractional deviation of $\sin^2\theta$ that is
measured by the $Z$-pole asymmetries, and $\zeta$ ($\le0$) is the
fractional deviation of the coupling of the $Z$ to ${j_L^3}$. In
deriving Eqs.~(\ref{lncsimple})--(\ref{szdef}) we have used the
results for $\delta M_W,$ $\delta\bar{g}$ and $\delta k$ given in
Eq.~(\ref{shifts}).

\subsection{Neutral-Current Interactions at Low Energy}

Experimental constraints on the effective value of $\sin^2\theta$
measured at low energy, via neutrino scattering and atomic parity
violation, no longer match the precision of measurements at high
energy which constrain $\delta_Z,$ $\zeta$ and $M_W,$ and thus will
not be part of our constraint analysis, which will be the subject of
the next Section.  Nevertheless, such a low-energy measurement of
$\sin^2\theta$ can in principle have different sensitivity to $Z'$
bosons, and so we conclude this Section by presenting the correction
to the low-energy value of $\sin^2\theta$ that would be measured in
this model. We can use the results for ${\cal L}_{\rm NC}$ to
calculate the the neutral-current matrix element, ${\cal M}_{\rm
NC}$, at zero momentum transfer. To establish notation, we first
mention that in the Standard Model, ${\cal M}_{\rm NC}$ is given (at
tree level) by
\begin{equation}
{\cal M}_{\rm NC}^0(q^2\approx0)={e^2\over q^2}Q\cdot Q'
-4\sqrt{2}G_F (I_3-s_0^2Q)\cdot(I_3'-s_0^2Q')\ ,
\end{equation}
where $(I_3,Q)$ and $(I_3',Q')$ stand for the matrix elements of the
neutral SU(2)$_W$ and electromagnetic currents in the external
fermionic states. The matrix element including corrections resulting
from the $W'$ isotriplet, ${\cal M}_{\rm NC},$ is then given at zero
momentum transfer by
\begin{eqnarray}
{\cal M}_{\rm NC}(q^2\approx0)&\approx&{e^2\over q^2}Q\cdot Q'\nonumber\\
&&-4\sqrt{2}G_F\bigg[(I_3-s_0^2(1+\delta_0)Q)
   \cdot(I_3'-s_0^2(1+\delta_0)Q')
\nonumber\\
& &+\left(2\delta_0+\mu(\kappa-\gamma)^2\right)s^2Q\cdot s^2Q'\bigg]\ .
\end{eqnarray}
Here $\delta_0$ is the $q^2=0$ analog of $\delta_Z$:
\begin{equation}
\delta_0={\delta\sin^2\theta\over\sin^2\theta}\Bigg|_{q^2\approx0}
  =\delta_Z
  +\mu(\kappa-\gamma)\left(\gamma+{s^2\over c^2-\mu}\kappa\right)\ .
\end{equation}
The correction to the coefficient of $Q\cdot Q'$ is unobservable in
practice. Note the absence of a correction to the term proportional
to $I_3\cdot I_3'$ in ${\cal M}_{\rm NC},$ i.e., $\rho(q^2=0)$ is
exactly $1$ (apart from the usual Standard-Model radiative
corrections), which is due to the custodial symmetry and the
constraint on $G_F$ measured in low-energy charged-current
interactions.  This removes some of the possible sensitivity to $Z'$
bosons.

We see that corrections to the low-energy theory due to the $W'$
enter through 4 independent functions of the $W'$ parameters: $\delta
M_W/M_W$, $\zeta$, $\delta_Z$, and $\delta_0$.  Again it should be
noted that all corrections to Standard-Model predictions vanish if
any two of the $W'$ parameters $\kappa$, $\gamma$ and $\mu$ vanish.
As mentioned, the first three of these functions, $\delta M_W/M_W,$
$\zeta$ and $\delta_Z,$ are measured with much greater accuracy than
the last, $\delta_0.$  We therefore ignore $\delta_0$ in our
constraint analysis, which we now present.

\section{Summary of Corrections in Terms of S, T and U}

The above corrections to Standard-Model predictions which result from
adding an isotriplet of $W'$ bosons can be conveniently summarized by
the $S$, $T$ and $U$ parameters introduced by Peskin and Takeuchi
\cite{pt,gr}. At a fundamental level, $S$, $T$ and $U$ are defined to
measure oblique corrections to Standard-Model predictions, i.e.,
corrections due to non-Standard particles appearing in vacuum
polarization graphs for the photon, $W$ and $Z$. However, at a
practical level, $S,$ $T$ and $U$ simply parameterize corrections to
the three Standard-Model quantities that are measured with high
precision (putting aside $\alpha$, $G_F$ and $M_Z$, which are fixed):
$M_W$, the coupling of the $Z$ to $j_L^3$, and $\sin^2\theta$
measured at the $Z$ pole. Therefore, although corrections due to the
$W'$ are in general nonoblique, by comparing the corrections to these
three quantities due to the $W'$ with their expressions in terms of
$S$, $T$ and $U,$ we can find the effective contributions of the $W'$
to $S,$ $T$ and $U.$

Contributions to $\delta M_W/M_W,$ $\delta_Z$ and $\zeta$ are given
in terms of $S$, $T$ and $U$ as \cite{pt},
\begin{eqnarray}
{\delta M_W\over M_W}&=&{1\over2}{\alpha\over c^2-s^2}
  \left[-{1\over2}S+c^2 T+{c^2-s^2\over4s^2}U\right]\nonumber\ ,\\
\delta_Z&=&{\alpha\over c^2-s^2}
  \left[{1\over4s^2}S-c^2 T\right]\nonumber\ ,\\
\zeta&=&\alpha T\ .
\label{STUdefs}
\end{eqnarray}
Equating these expressions with the corresponding expressions for
these quantities in terms of the $W'$ parameters, Eqs.~(\ref{shifts})
and (\ref{szdef}), we obtain the contributions of the $W'$ to $S$,
$T$ and $U$:
\begin{eqnarray}
{\alpha\over4c^2}S&=&-(1-\mu)\tilde{\kappa}(\mu\tilde{\kappa}+\gamma)
   \ ,\nonumber\\
\alpha T'&=&-\mu(\tilde{\kappa}+\gamma)^2\ ,\nonumber\\
{\alpha\over4s^2}U&=&\mu(\mu\tilde{\kappa}^2+2\tilde{\kappa}\gamma
   +\gamma^2)\ .
\label{stu}
\end{eqnarray}
Here we have introduced $\tilde{\kappa}\equiv s^2\kappa/(c^2-\mu),$
in terms of which the contributions of the $W'$ to $S,$ $T$ and $U$
can be expressed very concisely. We denote the contribution of the
$W'$ to $T$ as $T'$ because there is another important contribution
to $T,$ $T_{\rm top},$ due to the top quark. Using these expressions,
limits on $S,$ $T$ and $U$ can be converted into limits on the mass
and couplings of the $W'.$ However, the contribution to $T$ from the
top quark must first be removed, as we now describe.

\subsection{The Likelihood Function}

The electroweak constraints on $S$, $T$ and $U$ are combined
\cite{pt} by first constructing $\chi_0^2$:
\begin{equation}
\chi_0^2(S,T,U)=\sum_i\left[{x_i(S,T,U)-x_i^{\exp}\over\sigma_i}\right]^2
\ .
\end{equation}
Here the $x_i$ are electroweak observables, namely $M_W$, the $Z$
width,\footnote{In particular, we use the measurement of the leptonic
width of the $Z$, which is free of the theoretical uncertainties from
$\alpha_s$ and $\Gamma_b(Z)$ that plague the hadronic component of
the $Z$ width.} and $\sin^2\theta(M_Z)$, as shown in Table~1. The
$x_i(S,T,U)$ are the theoretical predictions for these observables
obtained by adding the oblique corrections linear in $S$, $T$ and $U$
to Standard-Model predictions; $x_i^{\exp}$ are the experimental
values; and $\sigma_i$ are the experimental errors. All these are
shown in Table~1.  The theoretical values are given for a $1000\,{\rm
GeV}$ Higgs with Standard-Model couplings.  Of course we do not know
the mass of the Higgs in the SCSM, though it should be of order the
weak scale. Further, unlike the Standard-Model Higgs, the couplings
of the SCSM Higgs to the $W$ bosons are unspecified.  However, this
represent a small uncertainty in the predictions of the SCSM which
does not alter our basic conclusions.

As mentioned above, there is a contribution from the top quark to
$T,$ which is quadratic in $m_t$ and can be sizable if the top is
heavy. (There are also small contributions from the top which are
only logarithmic in $m_t,$ which can safely be neglected.) In the
absence of information about the top mass, an arbitrarily negative
value of $T'$ could be canceled by an opposite, positive value of
$T_{\rm top}$ due to a heavy top. In this case, bounds on $T$ would
tell us nothing about $T',$ and only the bounds on $S$ and $U$ would
constrain the $W'$ parameters.

Of course, we now have information about the top mass from CDF, and
we use the result of their fit: $m_t=174\pm16\,$GeV \cite{top}.
However, the error in this measurement is not negligible, which can
be seen by noting that an $S$-$T$-$U$ analysis of the Standard Model
predicts the top mass with comparable uncertainty.  To incorporate
the CDF result for $m_t$, including the error, we convert it to a
measurement of $T_{\rm top}$: $T_{\rm top}=T_{\rm top}^0\pm\delta
T_{\rm top}.$  We then add a term to $\chi_0^2,$
\begin{equation}
\chi_t^2(S,T',U;T_{\rm top})=\chi_0^2(S,T'+T_{\rm top},U)+
  \left({T_{\rm top}-T_{\rm top}^0 \over\delta T_{\rm top}}\right)^2.
\end{equation}
This leads to a likelihood function $L_t(S,T',U;T_{\rm top})\equiv
N_t \exp[-\chi_t^2/2].$ Since we are here interested in the $W'$
parameters, and not $m_t,$ we integrate over $T_{\rm top}$ to find a
likelihood function of $S,$ $T'$ and $U$ alone:
\begin{equation}
L(S,T',U)=\int dT_{\rm top}\> L_t(S,T',U;T_{\rm top}) \equiv
N\exp[-\chi^2(S,T',U)/2]\ .
\label{mlf}
\end{equation}
In $L(S,T',U)$ the only unknowns are $\mu,$ $\kappa,$ and $\gamma,$
i.e., the $W'$ mass and couplings.  We will now exploit this
likelihood function to constrain these parameters.

\section{Constraints on a $W'$ in the SCSM}

\subsection{Bounds on $\kappa\gamma$}

{}From the expressions in Eq.~(\ref{stu}) for $S$, $T'$ and $U$ in
terms of $\kappa$, $\gamma$ and $\mu,$ we can derive bounds on the
product $\kappa\gamma.$ We first express the product
$\tilde{\kappa}\gamma$ as
\begin{equation}
\tilde{\kappa}\gamma=-{1 \over 1-\mu}{\alpha\over 4c^2}S
-\mu\tilde{\kappa}^2\ .
\end{equation}
Using $\alpha(T'+U/4s^2)=-\mu(1-\mu)\tilde{\kappa}^2,$ we find
\begin{equation}
\kappa\gamma={c^2-\mu \over 1-\mu}{\alpha \over s^2}
\left(-{S\over 4c^2} + T' + {U\over 4s^2}\right)\ .
\end{equation}
Then because $(c^2-\mu)/(1-\mu)$ is at most $c^2,$ $\kappa\gamma$ is
bounded as
\begin{equation}
{\alpha c^2 \over s^2}
\left(-{S\over 4 c^2} + T' + {U\over 4s^2}\right)_{\rm min}
< \kappa\gamma < - {\alpha c^2 \over s^2} S_{\rm min}\ ,
\label{kgbound}
\end{equation}
where in deriving the upper bound we used $\alpha(T'+U/4s^2)=
-\mu(1-\mu)\tilde{\kappa}^2<0.$ Here $S_{\rm min}$ refers to the
smallest (nonpositive) allowed value of $S.$ From the
$95\%$~Confidence Level (CL) bounds on $S$ and on $(-S/4c^2 + T' +
U/4s^2),$ obtained from the likelihood function in Eq.~(\ref{mlf}),
we find that
\begin{equation}
-0.049 < \kappa\gamma < 0.0055\qquad(95\%\ {\rm CL})\>.
\label{kgboundresult}
\end{equation}

\subsection{Allowed Region of $W'$ Mass and Couplings}

Our remaining results are obtained by exploring the volume of
$\kappa$-$\gamma$-$\mu$ space allowed by the likelihood function
(\ref{mlf}). Specifically, we consider points $(\kappa, \gamma, \mu)$
for which $(S, T', U)$ falls inside an $S$-$T'$-$U$ ellipsoid defined
by the value of $\chi^2$ corresponding to $95\%$~CL. This maximum
allowed value, $\chi^2_{\rm max},$ depends on the number of degrees
of freedom being constrained: for just one degree of freedom,
$\chi^2_{\rm max}=4$, which corresponds to two standard deviations,
while for two degrees of freedom, $\chi^2_{\rm max}\simeq6.18.$

A numerical search of the boundary of the allowed region of $(\kappa,
\gamma, \mu),$ defined by $\chi^2\leq4,$ shows that $\kappa\gamma$ is
bounded as
\begin{equation}
-0.028< \kappa\gamma < 0.0052\qquad(95\%\ {\rm CL})\> .
\label{kgboundnumerical}
\end{equation}
Hence we can state, with a confidence level of $95\%,$ that the $W$
boson must saturate the isovector electromagnetic form factor to
within $3\%.$

Figures 2 and 3 correspond to slices of the allowed region of
$(\kappa, \gamma, \mu)$ at fixed $\mu$ and $\gamma$ respectively.  In
Fig.~2 we show the allowed regions ($\chi^2_{\rm max}\le6.18$) of
$(\kappa, \gamma)$ for $M_{W'}=150$ and $400\,$GeV.  These
regions necessarily lie inside the hyperbolic bounds (solid lines)
which correspond to the extreme values of $\kappa\gamma$ allowed for
this value of $\chi^2_{\rm max}.$ Note that most of each allowed
region corresponds to both $\kappa$ and $\gamma$ much smaller than
one. However, when either $\kappa$ or $\gamma$ is extremely small,
the other can become large, particularly when the $W'$ is heavy.
This is expected since, as was pointed out in Section~3.2, if any two
of $\kappa,$ $\gamma$ and $\mu$ vanish, the remaining quantity is
unconstrained.

In Figure~3 we show the allowed range of $\kappa$ ($\chi^2\le4$) as a
function of $M_W/M_{W'}$ for $\gamma=1,$ $1/2,$ $1/3$ and $1/4.$ Here
we exploit the symmetry noted after Eq.~(\ref{lem'}) --- namely
invariance under simultaneous change of sign of $\gamma$ and $\kappa$
--- in order to restrict attention to positive $\gamma.$ From this
Figure we conclude that for reasonable values of $\gamma,$ $\kappa$
must be extremely small; in other words the $W'$ must mix much more
weakly with the photon than the $W$.  In particular, for
$\bar{g}'/\bar{g}>1/4$, $\kappa<0.025.$

In Figure~4 we show the maximum value of $\bar{g}'$ allowed by our
constraints ($\chi^2\le4$) as a function of $M_{W'},$  and we compare
our constraints with those obtained in the direct $W'$ search at CDF
\cite{W'search}. Note that our bounds are more restrictive than the
CDF bounds. Further, the CDF analysis assumes that the $W'$ decays
entirely into left-handed fermions, whereas in the SCSM a $W'$ will
primarily decay into $WZ$ for $M_{W'}$ above the decay threshold,
$M_W+M_Z$.  Hence the CDF bounds do not help us bound $\gamma$ in the
SCSM.  Our constraint analysis indicates that at a moderate value of
$M_{W'}$ such as $300\,{\rm GeV}$, the $W'$ coupling to fermions must
be less than a quarter of the $W$ coupling.

\section{Conclusion}

Comparing our results with the earlier analysis by Korpa and Ryzak
\cite{kr}, we can see how the continually improving electroweak
measurements have drastically pared away the allowed parameter space
in this model.  In their Figure~4 they found $\kappa\gamma$ allowed
to be as large as $0.2,$ compared to our upper bound of $0.0052.$
Similarly, for $\gamma=1$ they found that the $W'$ could be as light
as $170\,$GeV and $\kappa$ could be as large as $0.13,$ while for
$\gamma=1$ we now find that the $W'$ must be heavier than
$1075\,$GeV and $\kappa$ can be at most $0.006.$ Moreover, our much
more restrictive bounds hold at $95\%$ CL, while the earlier
bounds held only at $68\%$ CL.

Korpa and Ryzak concluded from their analysis that there was plenty
of room for the non-Standard particle content predicted by the SCSM.
{}From our analysis of the SCSM exploiting recent electroweak data, we
conclude that the currently allowed parameter space is so small as to
strongly argue against the model.  There is no reason to expect
vector dominance to hold at a level of $3\%.$ Nor can we understand
how the $W'$ could mix with the photon only $1/40$ as much as the $W$
mixes; yet we have found that this would have to be the case even if
the coupling of the $W'$ to the left-handed fermions is allowed to be
as small as $1/4$ the $W$ coupling (itself already small for a
strongly-coupled theory).

It is possible that by including more resonances in our analysis we
could find regions in the enlarged parameter space where the various
corrections to Standard-Model predictions cancel, without forcing the
masses and couplings of the resonances to be unnaturally small. But
from our analysis it is clear that these cancellations would have to
be rather delicate, and the agreement of the SCSM with experiment
would be just as inexplicable.

Of course, we can never completely exclude the SCSM solely on the
basis of experimental constraints.  The strongly-coupled dynamics
underlying the effective theory do not allow us to find predictions
for masses and couplings which could be contradicted by experiment.
However, the model offers no natural understanding of how it could
continue to evaded detection, disguised as the spontaneously-broken
Standard Model.  For this reason, we conclude that unless there
emerges from a study of the strong dynamics an explanation of how it
could be so nearly indistinguishable from the Standard Model, the
Strongly-Coupled Standard Model can no longer be viewed as a possible
theory of the electroweak interactions.

\section*{Acknowledgements}

We thank Lisa Randall for suggesting this research and Michael Peskin
for suggesting the $S$-$T$-$U$ analysis and for comments on the
manuscript. We also thank R.~L.~Jaffe and E.~Farhi for helpful
discussions.

\section*{Figure Captions}

\begin{itemize}
 \item[Figure 1.] Diagrammatic expansion of the isovector
electromagnetic form factor including the direct-coupling graph and
the $W$- and $W'$-pole graphs.
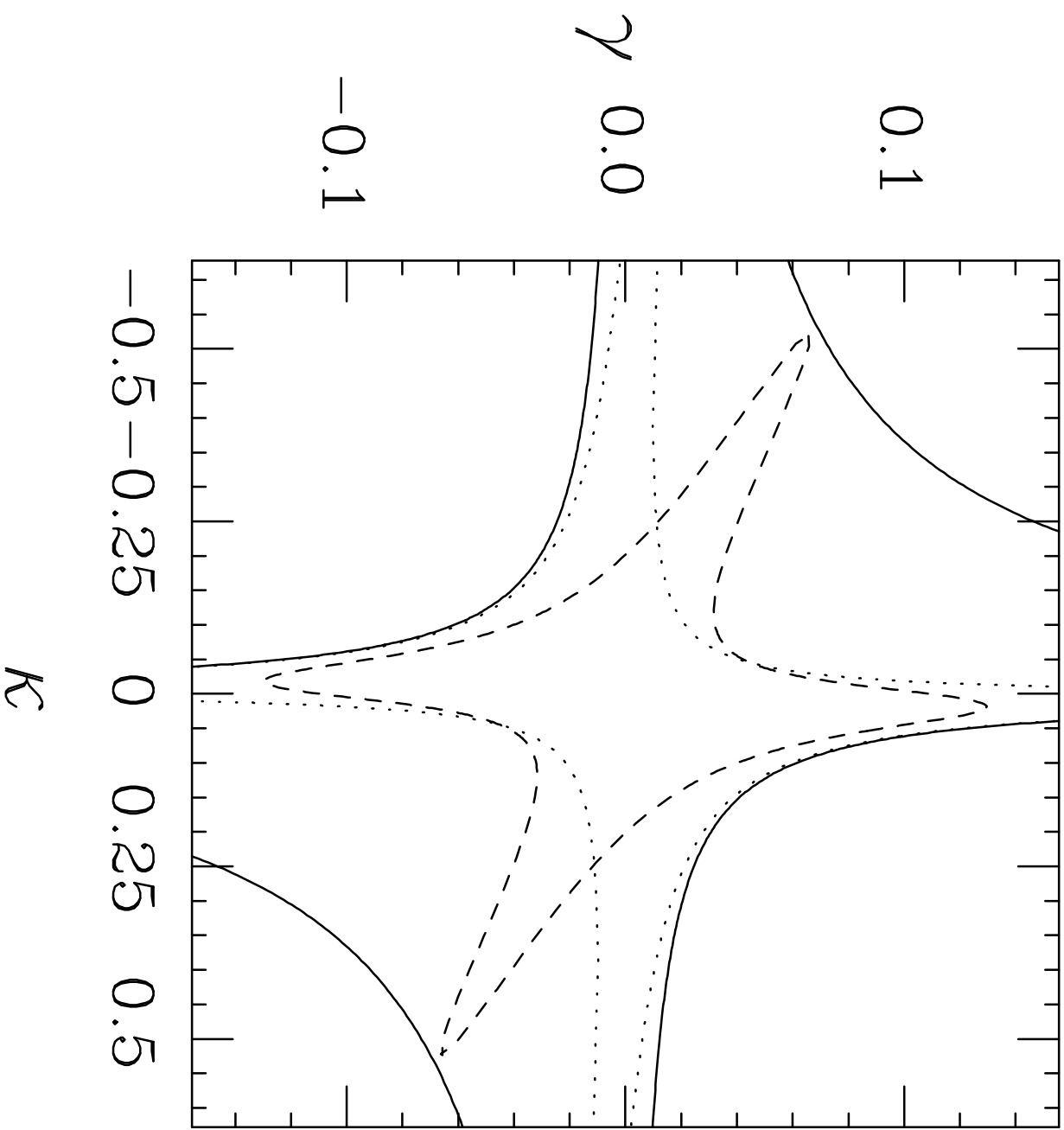
 \item[Figure 2.] Contours bounding the regions of the
$\kappa$-$\gamma$ plane allowed at $95\%$ CL for ${M_{W'}}=150\,$GeV
(dashes) and $400\,$GeV (dots). The solid lines are hyperbolas
defined by the bound on $\kappa\gamma$ at the same value of $\chi^2.$
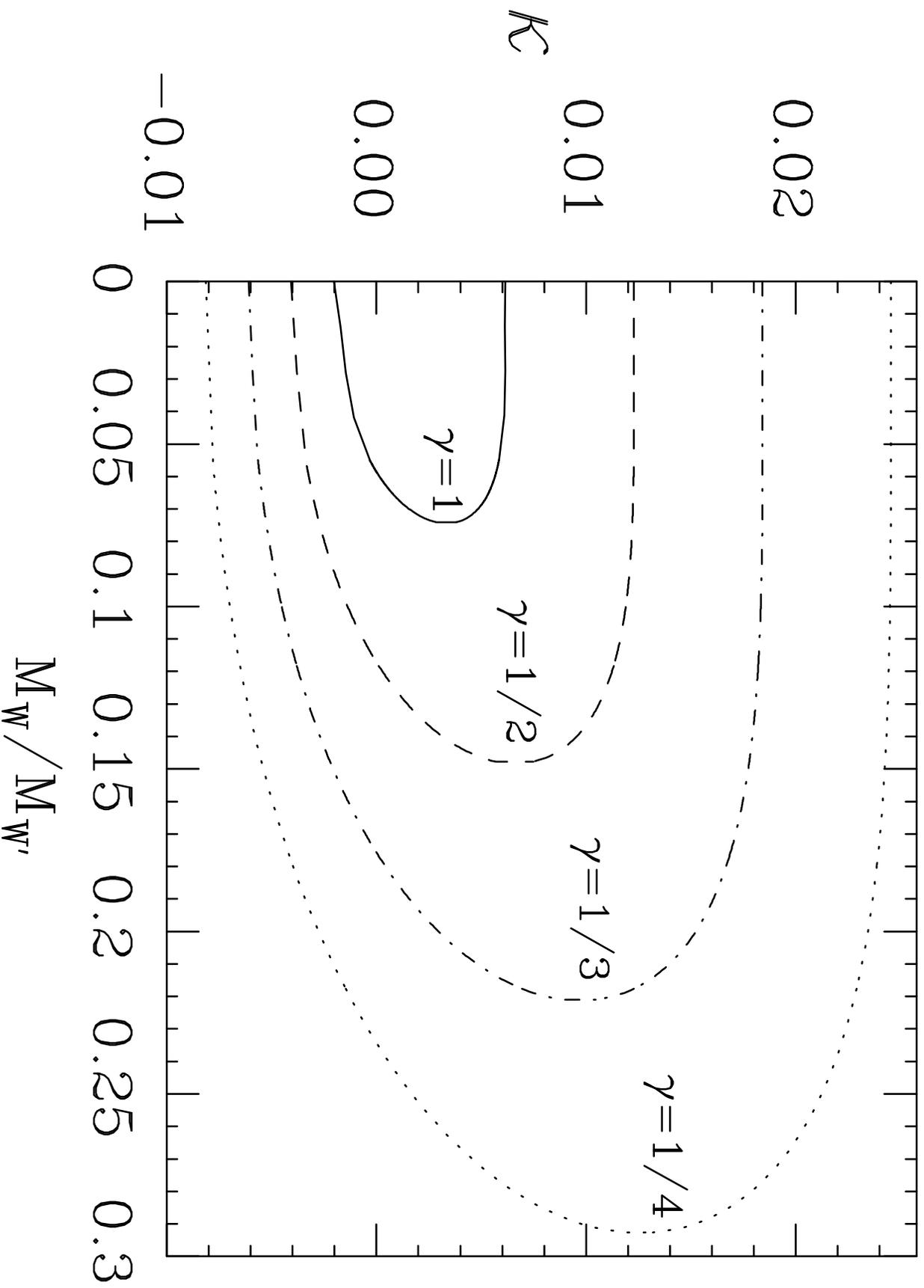
 \item[Figure 3.] Contours bounding the allowed region of the
$\kappa$-$M_W/{M_{W'}}$ plane for $\gamma=1$ (solid), $\gamma=1/2$
(dashes), $\gamma=1/3$ (dotdash) and $\gamma=1/4$ (dots).
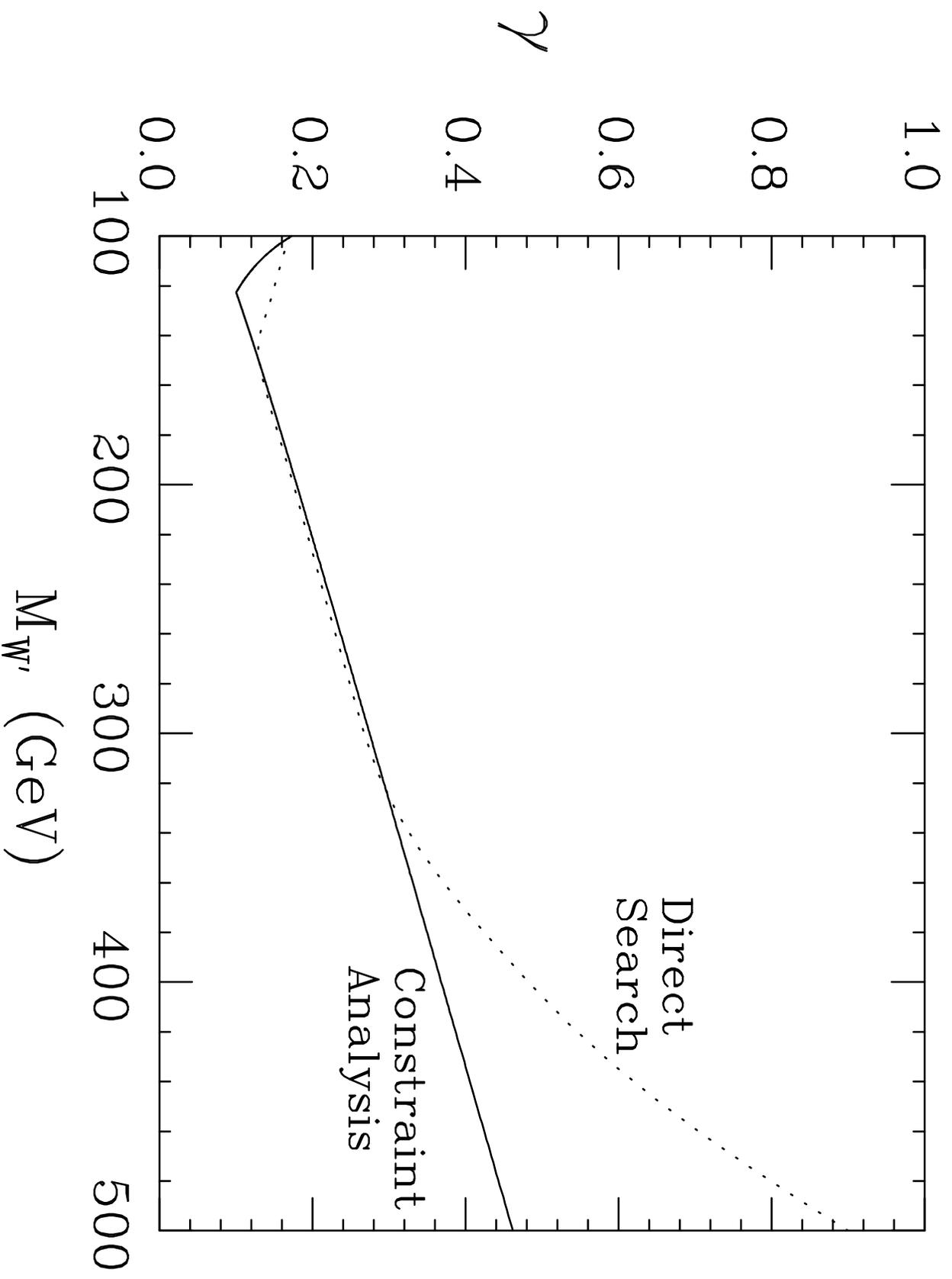
 \item[Figure 4.] Bounds on $\gamma$ ($95\%$ CL) for a range of
$M_{W'}.$  The dotted curve interpolates through the bounds obtained in
the direct $W'$ search at CDF \cite{W'search} while the solid curve
shows the bounds obtained in our analysis of electroweak constraints
applied to the SCSM.
 \end{itemize}

\section*{Table Caption}

\begin{itemize}
 \item[Table 1.] Theoretical and measured values of the electroweak
observables used to constrain the $W'$ couplings.  The theoretical
values correspond to vanishing $W'$ couplings, a top mass of
$174\,{\rm GeV}$, and a Higgs of mass $1000\,{\rm GeV}$
(with Standard-Model couplings).
 \end{itemize}

\section*{Table~1}

\begin{center}
\begin{tabular}{|l|l|l|l|}\hline
 Observable  & Theoretical Value  & Measured Value  & Experiment\\ \hline
 $M_W$       & $80.23\,$GeV       & $80.23\pm0.18\,$GeV
    & CDF and D0 \cite{Wmass}; UA2 \cite{ua2}  \\ \hline
 $\Gamma(Z\to{\rm leptons})$
             & $83.68\,$MeV       & $83.96\pm0.18\,$MeV
    & LEP \cite{leptwidth}                     \\ \hline
 $\sin^2\theta(M_Z)$ & $0.2331$           & $0.2317\pm0.0004$
    & LEP \cite{leps2}; SLD \cite{sld}         \\ \hline
\end{tabular}
\end{center}


\begin{thebibliography}{9}
 %
\bibitem{af} L.~F.~Abbott and E.~Farhi, Phys.\ Lett.\ B 101
  (1981) 69; Nucl.\ Phys.\ B 189 (1981) 547.
\bibitem{kr} C.~Korpa and Z.~Ryzak, Phys.\ Rev.\  D
  34 (1986) 2139.
\bibitem{hs} P.~Q.~Hung and J.~J.~Sakurai, Nucl.\ Phys.\ B 143
  (1978) 81.
\bibitem{cfj} M.~Claudson, E.~Farhi, and R.~L.~Jaffe,
  Phys.\ Rev.\  D 34 (1986) 873.
\bibitem{comp} T.~Banks and E.~Rabinovici, Nucl.\ Phys.\ B 160 (1979)
  349;\\ E.~Fradkin and S.~Schenker, Phys. Rev. D 19 (1979) 3682;\\
  G.~'t~Hooft, in {\it Recent Developments in Gauge Theory,} edited by
  G.~'t~Hooft et al. (Plenum, New York, 1980);\\
  S.~Dimopoulos, S.~Raby and L.~Susskind, Nucl.\ Phys.\ B 173 (1980)
  208.
\bibitem{pt} M.~E.~Peskin and T.~Takeuchi, Phys.\ Rev.\  D 46
  (1992) 381.
\bibitem{gr} Similar parameters were discussed in M.~Golden and
  L.~Randall, Nucl.\ Phys.\ B 361 (1991) 3.
\bibitem{top} CDF Collaboration, F.~Abe et al., FERMILAB-PUB-94-097-E;
  Phys.\ Rev.\ Lett. 73 (1994) 225.
\bibitem{Wmass} M.~Demarteau et al., {\it Combining $W$ Mass
  Measurements}, CDF/PHYS/CDF\-/PUBLIC/2552 and D0NOTE 2115, May 1994.
\bibitem{ua2} UA2 Collaboration, J.~Alitti et al., Phys.\ Lett.
  B 276 (1992) 354.
\bibitem{leptwidth}Internal note of the Subgroup of the LEP Electroweak
  Working Group on $Z^0$ Lineshape and Lepton Forward-Backward
  Asymmetries, P.~Clarke et al., {\it Updated Parameters of the $Z^0$
  Lineshape and Lepton Forward-Backward Asymmetries from Combined
  Preliminary Data of the LEP Experiments}, LEPLINE 94-01, ALEPH 94-120
  PHYSIC 94-104, DELPHI 94-99 PHYS 416, L3 Note 1629, OPAL Technical
  Note TN244, July 1994.
\bibitem{leps2} Internal note of the LEP Electroweak Working Group,
  A.~Blondel et al., {\it Constraints on Standard Model Parameters from
  Combined Preliminary Data of the LEP Experiments}, LEPEWWG/94-02,
  ALEPH 94-121 PHYSIC 94-105, DELPHI 94-23 Phys 357, L3 note 1577,
  OPAL Technical Note TN245, July 1994.
\bibitem{sld} SLD Collaboration, K.~Abe et al., Phys.\ Rev.\
  Lett. 73 (1994) 25.
\bibitem{W'search} CDF Collaboration, F.~Abe et al., Phys.\ Rev.\
  Lett. 67 (1991) 2609.
\end{thebibliography}
\end{document}